# High coercivity induced by mechanical milling in cobalt ferrite powders

Ponce, A. S.[1], Chagas, E. F.[1], Prado, R. J.[1], Fernandes, C. H. M.[2], Terezo, A. J.[2], and Baggio-Saitovitch, E.[3]

[1]*Instituto de Física, Universidade Federal de Mato Grosso, 78060-900, Cuiabá-MT, Brazil*

[2]*Departamento de Química, Universidade Federal do Mato Grosso, 78060-900, Cuiabá-MT, Brazil*

[3]*Centro Brasileiro de Pesquisas Físicas, Rua Xavier Sigaud 150 Urca. Rio de Janeiro, Brazil.*

Phone number: 55 65 3615 8747

Fax: 55 65 3615 8730

Email address: efchagas@fisica.ufmt.br

## Abstract

In this work we report a study of the magnetic behavior of ferrimagnetic oxide $CoFe_2O_4$ treated by mechanical milling with different grinding balls. The cobalt ferrite nanoparticles were prepared using a simple hydrothermal method and annealed at 500°C. The non-milled sample presented coercivity of about 1.9 kOe, saturation magnetization of 69.5 emu/g, and a remanence ratio of 0.42. After milling, two samples attained coercivity of 4.2 and 4.1 kOe, and saturation magnetization of 67.0 and 71.4 emu/g respectively. The remanence ratio $M_R/M_S$ for these samples increase to 0.49 and 0.51, respectively. To investigate the influence of the microstructure on the magnetic behavior of these samples, we used X-ray powder diffraction (XPD), transmission electron microscopy (TEM), and vibrating sample magnetometry (VSM). The XPD analysis by the Williamson-Hall plot was used to estimate the average crystallite size and strain induced by mechanical milling in the samples.

## Introduction

Cobalt ferrite ($CoFe_2O_4$) is a promising compound with potential biomedical applications

[1,2], as well as use in high density magnetic storage [3], permanent magnets [4], and electronic devices. This is due to several properties, such as electrical insulation, chemical stability, high magnetic-elastic effect [5], moderate saturation magnetization, and high coercivity ($H_C$).

Particularly in permanent magnet applications, high $H_C$ is a fundamental characteristic mainly because $H_C$ is an important parameter to the maximum energy product $(BH)_{max}$, the figure of merit for permanent magnets. Many authors report tuning the HC through thermal annealing [6], capping [7] and mechanical milling treatment [8] of the grains. Limaye *et al.* [7] obtained cobalt ferrite nanoparticles with very high coercivity (9.5 kOe) by capping it with oleic acid; however, the saturation magnetization decreased to 7.1 emu/g – that is, about 10 times less than that obtained from uncapped nanoparticles. This small value of saturation magnetization ($M_S$) is an undesirable property in hard magnetic applications. In another interesting article, Liu *et al.* [9] increased the $H_C$ of cobalt ferrite, from 1.23 to 5.1 kOe, with a relatively small decrease in $M_S$. They used a brief (1.5 h) mechanical milling process on relatively large particles (average grain size of 240 nm). However, for nanoparticles with an average grain size of 12 nm, no improvement in magnetic parameters was obtained.

Application of mechanical milling to nanoparticles (diameter < 100 nm) seeking to increase $H_C$ has not yet been successful. However, few important phenomena happen only at the nanoscale. Exchange spring (exchange coupled) is an example. High $H_C$ cobalt ferrite make possible an expressive increase in the $(BH)_{max}$ in exchange coupled nanocomposite $CoFe_2O_4/CoFe_2$ [10,11]. The importance of studying the magnetic behavior of this nanocomposite, aiming to improve its magnetic properties, is what led to this work.

In order to increase the $H_C$ of cobalt ferrite nanoparticles, we processed them with thermal annealing at moderate temperature (500 °C) and brief mechanical milling. Using a shake miller and several different milling parameters, we increased the $H_C$ of all treated samples. The result of this processing was an increase of about 150% in coercivity. Also, the $M_R/M_S$ ratio improved from 0.42 to 0.49 and 0.51 for the two best samples, and the $(BH)_{max}$ was amplified almost 100%. This substantial enhancement in magnetic parameters justifies the publication of this study. Samples were analyzed by X-ray powder diffraction (XPD), transmission electron microscopy (TEM), and magnetic hysteresis curves, all at room temperature.

**Experimental procedure**

The cobalt ferrite nanoparticles were synthesized using a simple hydrothermal method described in a previous work [10]. The morphology and particles size distribution of the samples were examined by direct observation via transmission electron microscopy (TEM) using JEOL-2100 apparatus installed at LNNano / LNLS – Campinas – Brazil woking at 200 kV.

The pristine sample (without thermal neither mechanical milling treatments) was annealed at 500 °C for 6 hours and separated in 5 amounts called CF0, CF1, CF2, CF3 and CF4. The sample CF0 was not milled and the other samples were milled during 1.5 hours using different milling parameters. The milling process used was the high energy mechanical ball milling in a Spex 8000 miller. Details of the different milling processes to each sample are described in table I.

Table I – Details of mechanical milling process.

| Sample | Ball material (mass density) | Ball diameter | Proportion Ball/sample | Milling time (h) |
|---|---|---|---|---|
| CF1 | Stainless steel (7.8 g/cm$^3$) | 6.5 mm | 1:4 | 1.5 |
| CF2 | Zirconia (6.1 g/cm$^3$) | 6.0 and 1.2-1.4 mm* | 1:7:3 | 1.5 |
| CF3 | Zirconia (6.1 g/cm$^3$) | 6.0 mm | 1:7 | 1.5 |
| CF4 | Tungsten carbide (15 g/cm$^3$) | 5.0 mm | 1:9 | 1.5 |

* The sample CF2 was milled by balls of both indicated sizes.

The crystalline phases of the cobalt ferrite nanoparticles were identified by X-ray powder diffraction (XPD) patterns, obtained on a Shimadzu XRD-6000 diffractometer installed at the "*Laboratório Multiusuário de Técnicas Analíticas*" (LAMUTA / UFMT – Cuiabá-MT - Brazil), equipped with graphite monochromator and conventional Cu tube (0.154178 nm) working at 1.2 kW (40 kV, 30 mA), using the Bragg-Brentano geometry.

Magnetic measurements were carried out using a vibrating sample magnetometer (VSM) model VersaLab Quantum Design installed at CBPF, Rio de Janeiro-RJ – Brazil. Experiments were done at room temperature and using magnetic field up to 2.5 T.

**Results and discussion**

The XPD patterns obtained from samples CF0, CF1, CF2, CF3 and CF4 (see figure 1) confirming that all samples are $CoFe_2O_4$ with the expected inverse spinel structure. These measurements indicate the absence of any other phases or contamination before and after milling processing.

TEM images of the pristine cobalt ferrite nanoparticles revealed an extremely polydisperse system with several forms (see figure 2*a*). The particle size distribution indicates that cobalt ferrite particles have a mean diameter of 22.6 nm; however TEM images also revealed the presence of particles larger than 100 nm. TEM measurements also indicate that nanoparticles increased after thermal treatment, changing the mean diameter to 44.7 nm. As occurred for the pristine sample, some big particles were observed to sample CF0 (figure 2*b*). These results indicate that, despite the relatively low annealing temperature, there was coalescence of nanoparticles causing the increase of mean size.

For samples CF3 and CF4 the TEM images revealed similar particles with shear bands (see the figure 3*a* and 3*b)*. We observed both regular and irregular shear bands in good agreement with the microstructural evolution described by Liu *et al*. [9]. These observations suggest that the shear bands are moiré fringes (see figure 3*c*).

To evaluate the influence of mechanical milling on the magnetic behavior of our samples, we measured the magnetic hysteresis loop. These measurements are shown in figure 4. Three interesting characteristics can be noted from the hysteresis loops obtained from milled samples: the increase in $H_C$, the changes in $M_S$, and the increase of $M_R$ to all milled samples. These characteristics are discussed in detail below.

Samples CF3 and CF4 presented the highest $H_C$ values found in this work, 4.1 and 4.2 kOe, respectively. This enhancement in $H_C$, when compared with the value obtained for sample CF0 (no mechanical milling), is very expressive and indicates an increase of about 120% after milling, for both samples. However, the maximum energy product for sample CF4 (0.88 MGOe) was smaller than that observed in sample CF3 (1.1 MGOe) due the different $M_R$ for both samples – 36.8 emu/g for sample CF3 and 33.0 emu/g for CF4. The main magnetic characteristics obtained for all the samples studied in this work are shown in Table II.

Table II – Magnetic characteristics of milled cobalt ferrite at room temperature.

| Sample | Coercivity (kOe) | $M_R$(emu/g) | $M_r/M_s$ | $(BH)_{max}$ (MGOe) | $M_S$(emu/g) |
|---|---|---|---|---|---|
| CF0 | 1.9 | 29.2 | 0.42 | 0.54 | 69.5 |
| CF1 | 2.5 | 29.4 | 0.44 | 0.61 | 66.6 |
| CF2 | 3.7 | 29.3 | 0.45 | 0.66 | 64.6 |
| CF3 | 4.1 | 36.8 | 0.51 | 1.1 | 71.4 |
| CF4 | 4.2 | 33.0 | 0.49 | 0.88 | 67.0 |

To investigate the influence of the annealing and milling processes on the strain and average crystallite size of the samples, we used the Williamson-Hall plot [12-14] obtained by the XPD measurements, as in reference [13]. The analysis shows that broadening of the diffraction peaks after milling evidences both grain size reduction and increase in microstructural strain. The instrumental broadening was corrected using the $Y_2O_3$ diffraction pattern as the standard. The results of the Williamson-Hall plot analysis are shown in figure 5.

To facilitate the analysis of the influence of structural parameters on the magnetic behavior of the samples, we included in Table III results obtained from the Williamson-Hall analysis together with coercivity and saturation magnetization.

Table III – Results obtained from Williamson-Hall analysis, coercivity and saturation magnetization.

| Sample | Grain size (nm) | Strain (%) | $H_C$ (kOe) | $M_S$ (emu/g) |
|---|---|---|---|---|
| CF0 | 64(5) | 0.32(3) | 1.9 | 69.5 |
| CF1 | 34(2) | 0.34(3) | 2.5 | 66.6 |
| CF2 | 17(2) | 0.8(1) | 3.7 | 64.6 |
| CF3 | 25(2) | 0.8(1) | 4.1 | 71.4 |
| CF4 | 21(1) | 1.0(1) | 4.2 | 67.0 |

As one can see, milled samples presented a smaller average crystallite size, and microstructural strain-induced by milling was observed in samples CF2, CF3 and CF4. The strain from samples CF0 and CF1 were similar, within the margin of experimental errors of the method. The strain observed in sample CF4 was very close to the value obtained by Liu and Ding [8] for their sample with highest coercivity. The main difference between our sample CF4 and that of Liu and Ding's [8] work is the average size of the particles. In Liu and Ding's work the average particle size was about 110 nm, while in our study it was 21 nm. Besides the mean diameter of the nanoparticles from sample CF4 being smaller than the critical size expected for the magnetic multi-domain regime [15], many big particles with a diameter larger than this critical size were observed in TEM images. Thus we conclude that particles in the multi-domain regime play an important role in magnetic behavior. Pinning effects observed by Liu *et al.* [9] must also be important to our samples. Structural defects that produce the pinning of wall domains were observed with high magnification TEM measurements (see figure 6). We conclude that a combination of strain and pinning effect is responsible for the increase of $H_C$ in sample CF4.

The increase of $H_C$ observed in the other samples (CF1, CF2, and CF3) is clearly associated with the strain induced by mechanical milling. The samples presenting the two highest $H_C$ values, CF3 ($H_C$ = 4.1 kOe) and CF4 ($H_C$ = 4.2 kOe), also have the largest strains: 0.8 and 1.0, respectively. The sample with lowest $H_C$, CF1 ($H_C$ = 2.5 kOe), presents the smaller strain (0.34). To sample CF3, as to sample CF4, shear bands were observed in TEM images (not showed) suggesting that pinning effect play an important role in the magnitude of $H_C$ also to this sample.

The second important characteristic observed in the hysteresis loops was the change in saturation magnetization ($M_S$) in the milled samples. The $M_S$ values were obtained by extrapolating 1/H to zero in the plot magnetization versus 1/H. Except for the sample CF3, the $M_S$ decreased in the other milled samples. $M_S$ increased from 69.5 emu/g (sample CF0) to 71.4 emu/g in the sample CF3, but for sample CF2 there was a decrease of about 7.5%. Normally, when $M_S$ decreases together with crystallite size, the decrease in $M_S$ is attributed to the surface effect, sometimes called the "dead" surface [16]. The dead surface is associated with disorder of surface spins. The number of surface spins in a sample increases when the crystallite size decreases, thus it is expected that the $M_S$ will also decrease. However, this explanation is not consistent with all our results. If compared with the non-milled sample CF0, the saturation magnetization of the samples CF1, CF2 and CF4 decreases with the crystallite size, but the $M_S$ for sample CF3 was higher than that obtained for sample CF0 (see Table III). Thus we attributed the decrease in saturation magnetization to a combination of two factors: the surface effect [17] and redistribution of magnetic cations [18].

To understand the cation redistribution it is necessary to know the cobalt ferrite structure and the distribution of magnetic ions in this structure. The cobalt ferrite has an inverse spinel structure with the two magnetic ions ($Co^{2+}$ and $Fe^{3+}$) occupying two different sites called A and B. The A site is a tetrahedral site and the B site is an octahedral site. In an ideal inverse spinel cobalt ferrite, half of the $Fe^{3+}$ cations (magnetic moment equal to $5\mu_B$) occupy the A-sites and the other half the B-sites, together with $Co^{2+}$ cations. (There are two B sites for formula unit.) Since the magnetic moments of the ions in the A and B sites are aligned in an anti-parallel way, the magnetic contribution of the $Fe^{3+}$ cations are mutually compensated. Therefore, the net magnetic moment of the ideal inverse spinel cobalt ferrite is due only to the Co2+ cations (magnetic moment equal to 3μB). However, the cobalt ferrite normally presents a mixed spinel structure, the sites A and B have a fraction of Co2+ and Fe3+ cations. Therefore, in a general way, the structure of cobalt ferrite can be described by reference [18]:

$$Co^{2+}_{1-i}Fe^{3+}_{i}[Co^{2+}_{i}Fe^{3+}_{2-i}]O^{2-}_{4} \qquad (1)$$

where *i* (the degree of inversion) describes the fraction of the tetrahedral sites occupied by $Fe^{3+}$ cations. The ideal inverse spinel structure has *i* = 1, *i* = 0 to the normal spinel and to the mixed spinel structure the value of *i* is between 0 and 1.

If $Co^{2+}$ ions move from A-sites to B-sites and consequently $Fe^{3+}$ cations move from B-sites to A-sites, there will be a change in the degree of inversion *i* (see equation 1), a decrease of the net magnetic moment of the material, and, consequently, a decrease in $M_S$. On the other hand, if the opposite process happens the saturation magnetization will increase. In a recent article, Sato Turtelli *et al.* [18] showed that samples prepared by mechanical milling present an *i* factor smaller than that obtained for samples synthesized by sol gel method. Another indication that there is a cationic redistribution in $CoFe_2O_4$, due mechanical milling is the work of Stewart *et al.* [19]. They observed cation swap between magnetic site A and B in normal spinel ferrite $ZnFe_2O_4$ due the mechanical milling. In addition, the cationic redistribution described before is consistent with the enhancement of the $H_C$ due to changes in the magneto-crystalline anisotropy. According to Sato Turtelli *et al.* [18] more $Co^{2+}$ in B-site (smaller value of *i* factor) increases the magneto-crystalline anisotropy and, consequently, increases the coercivity.

The third interesting result observed in magnetic measurements is the increase of remanence ($M_R$) and, consequently, the increase of the remanence ratio ($M_R/M_S$) obtained for all of the milled samples. Sample CF3 presented the most expressive increase of $M_R$ and, consequently, highest $M_R/M_S$ ratio (0.51). There are two possible causes of this effect (desirable for hard magnetic applications): first are the changes of magnetic domain regime between sample CF0 and samples CF1, CF2, CF3 and CF4. The sample CF0 has an average size of about 65 nm, indicating that this sample is in magnetic multi-domain regime [15]. Thus the magnetization process is caused mainly by the movement of domain walls. However, when the grain is smaller than 40 nm the cobalt ferrite is in single magnetic domain regime and the magnetization curve is governed by rotation of the magnetic moments. The second possible explanation is the enhancement of cubic magneto-crystalline anisotropy, indicated by the approximation of the theoretical value expected to non-interacting particles with positive cubic anisotropy (0.83 at 0K) [20]. Similar effect occurs when the temperature decreases [10,21].

**Conclusion**

In agreement with other authors, our results suggest that strain plays an important role in the $H_C$ value. Structural defects were observed in big particles (around of 100 nm), these pinning centers must be important to high $H_C$ behavior.

The increase of $M_S$ observed in sample CF3 suggest that high energy ball milling can promote a cationic swap between $Co^{2+}$ and $Fe^{3+}$ from A and B sites respectively. However, this assumption deserves more investigation.

We achieved a significant increase of the $(BH)_{max}$ for sample CF3 compared with non-milled samples due the $H_C$ and the $M_R$ enhancement. The increase of the $(BH)_{max}$ in sample CF4 was smaller, indicating that grinding ball mass density plays an important role in the magnetic behavior of milled samples.

**Acknowledgments**

This work was supported by CAPES Brazilian funding agency (PROCAD-NF project #2233/2008). The authors would like to thank the LNNano/LNLS for technical support during electron microscopy work

**Figures**

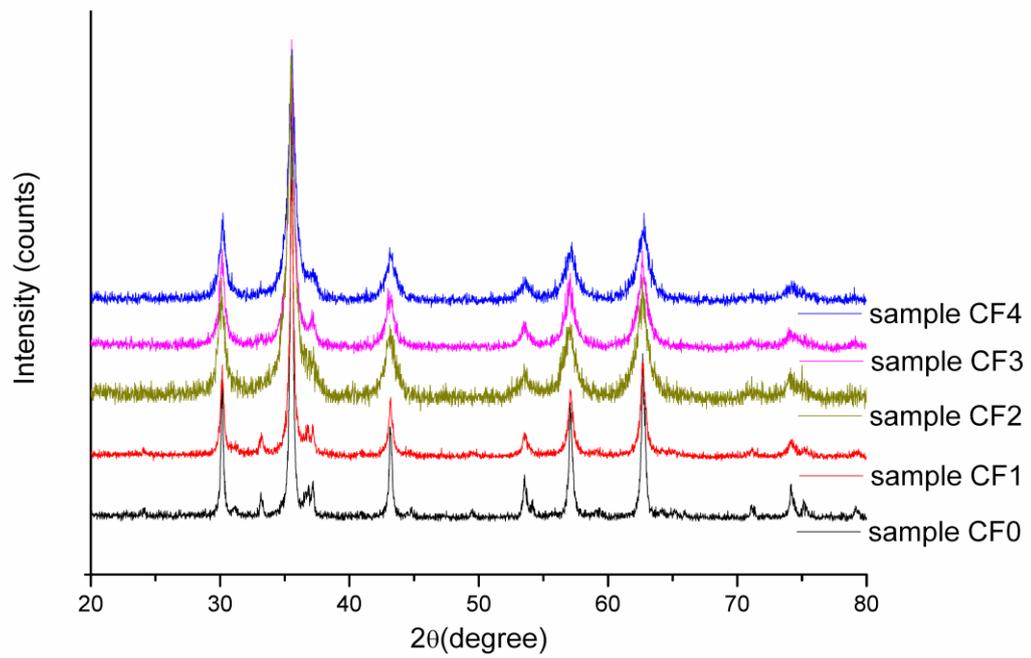

Figure 1 – XPD patterns of Co ferrite.

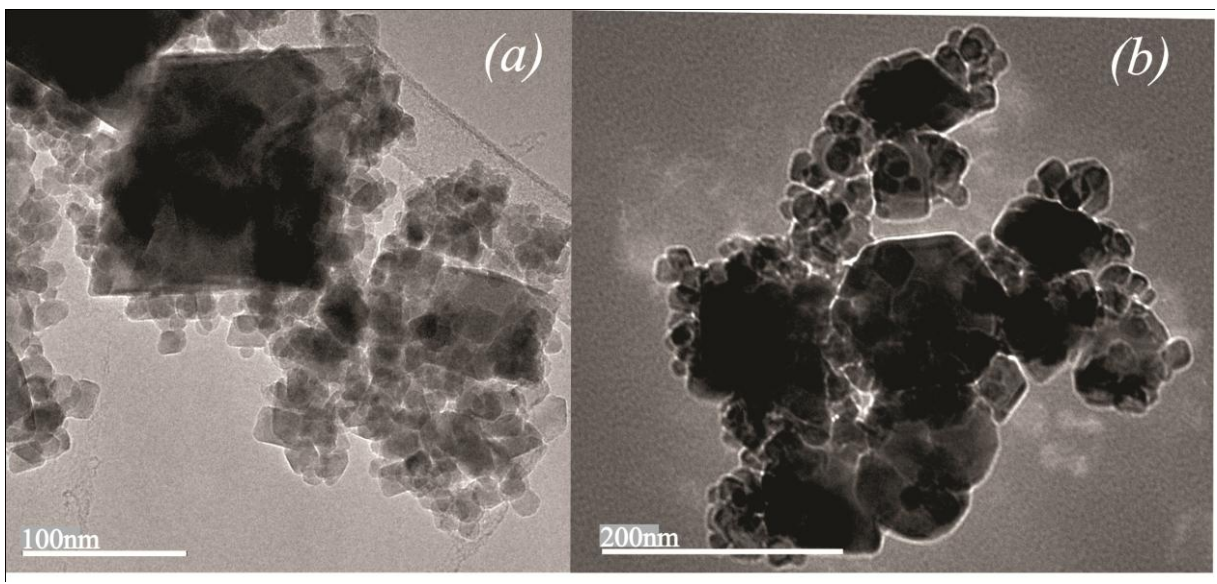

Figure 2 – TEM images of *a)* the pristine sample and *b)* sample CF0.

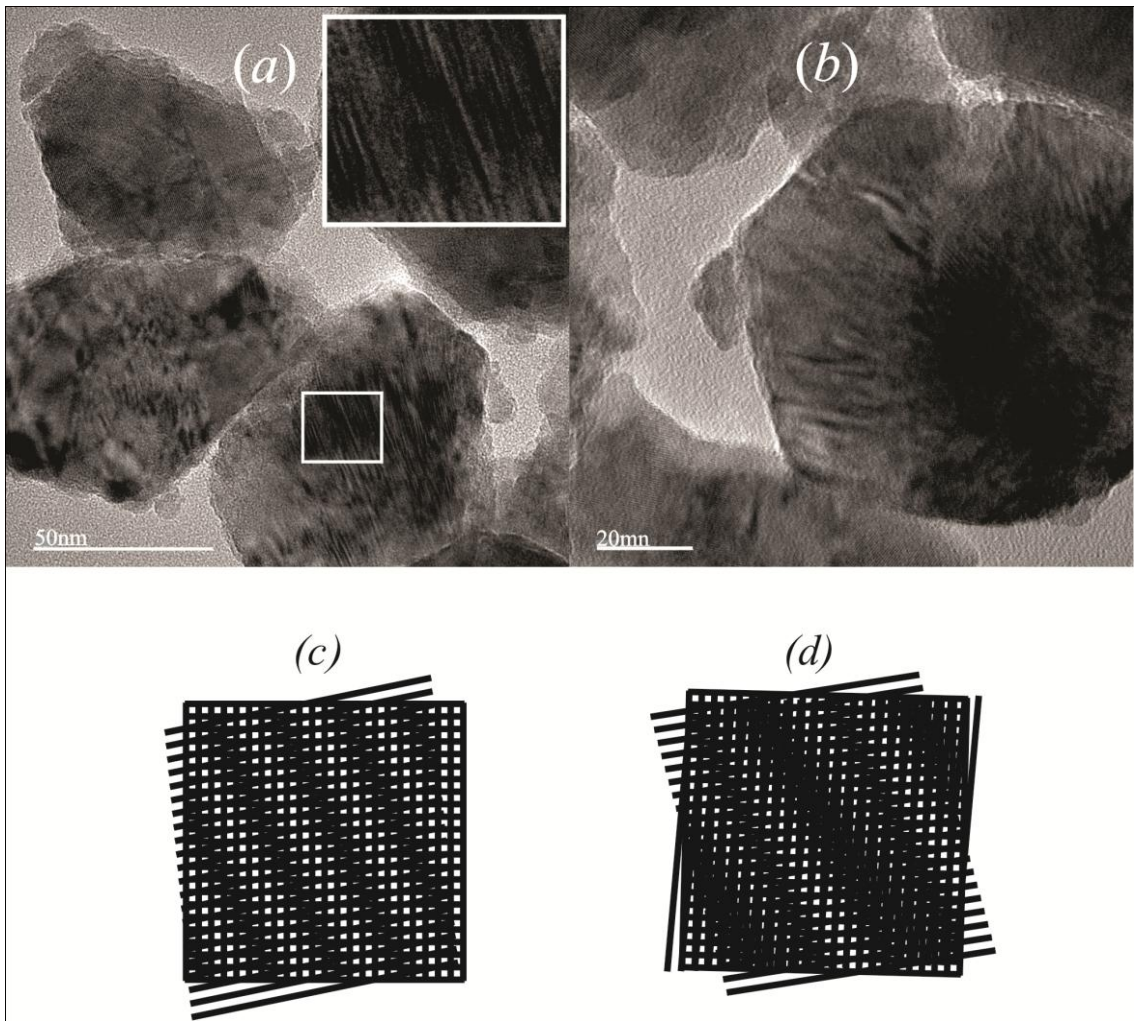

Figure 3 – TEM images showing shear bands to a) CF4 and b) CF3 samples TEM image, c) regular, and d) irregular squematic pictures of moire fringes. The insert of *a*) shows details of the marked area.

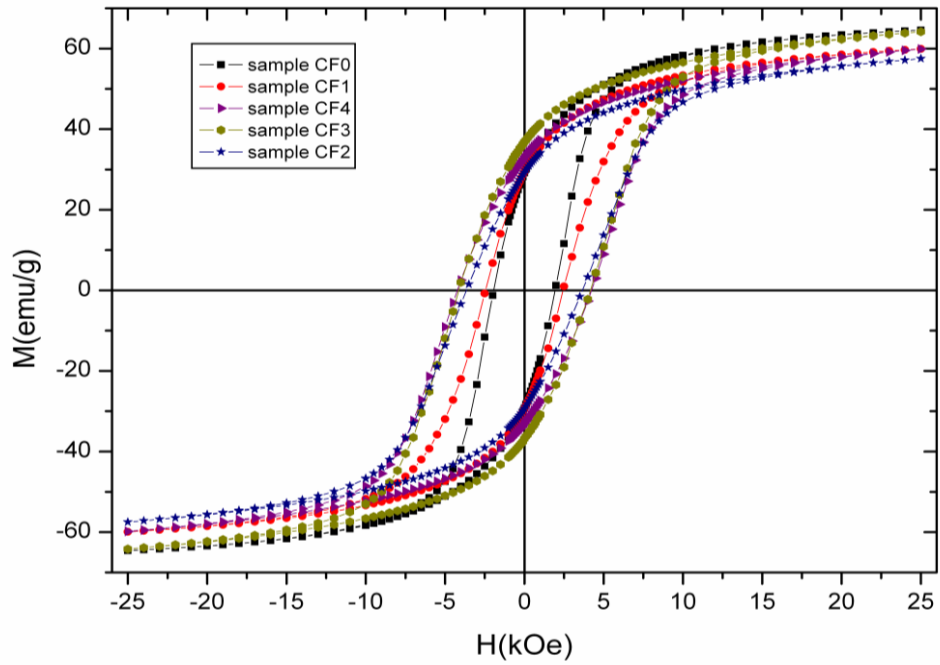

Figure 4 - Hysteresis loops for $CoFe_2O_4$ nanoparticles at room temperature (300 K) for maximum applied field of 25 kOe.

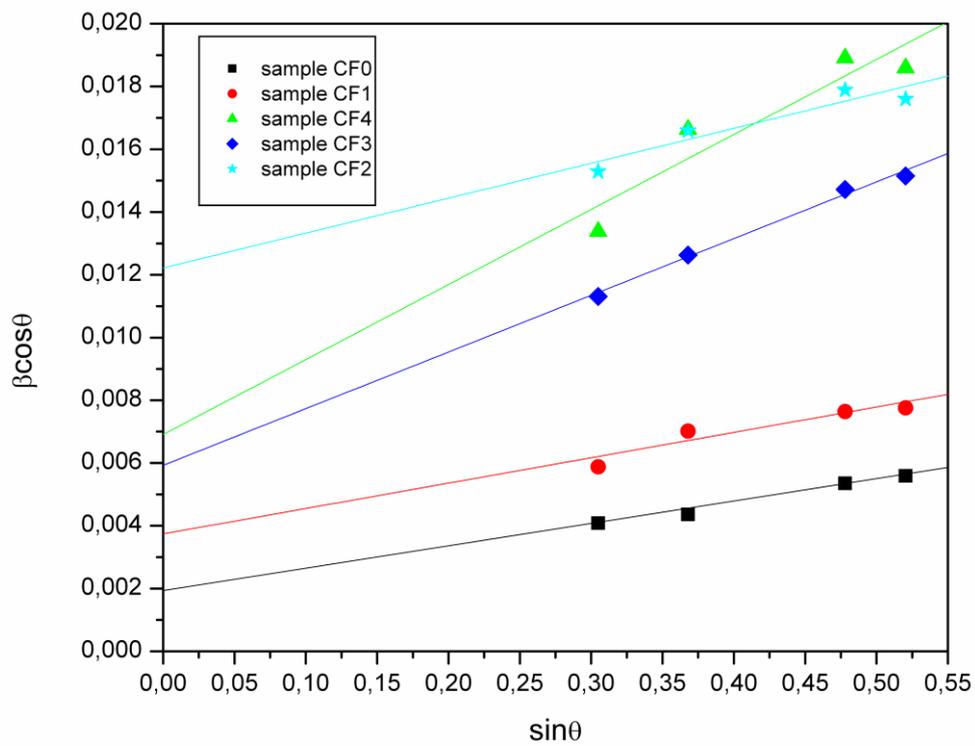

Figure 5 – Willianson-Hall plot to milled cobalt ferrite.

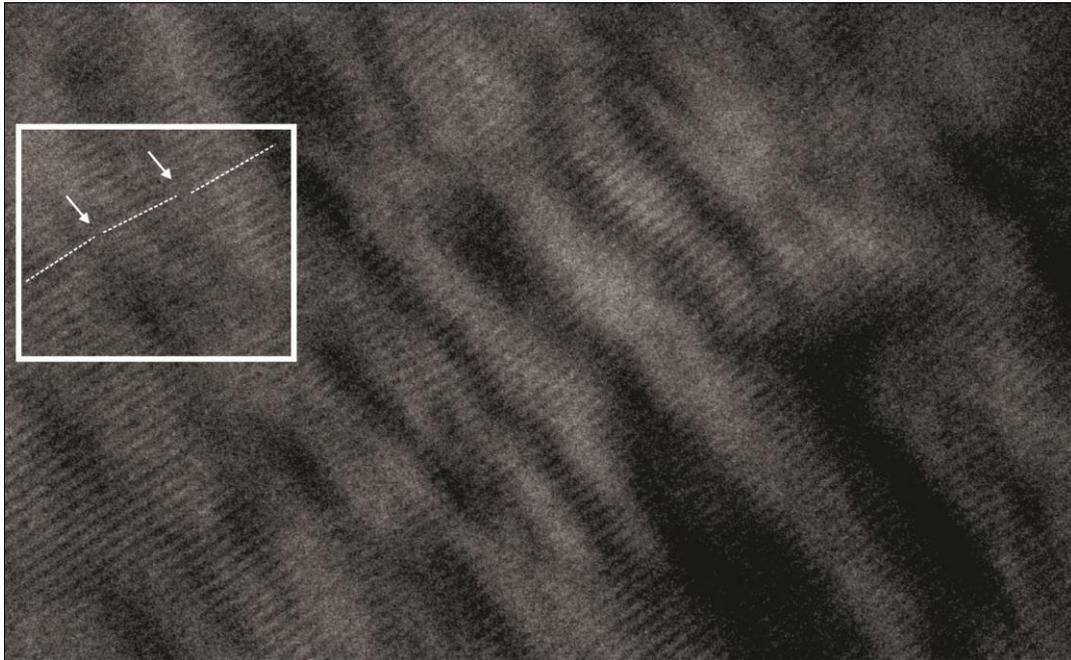
Figure 6 - Structural defects in crystalline plans to milled cobalt ferrite (sample CF4). The marked area indicate a structural defect.